\documentclass[twocolumn,letter]{jpsj2} 
\def\Vec#1{\mbox{\boldmath $#1$}}
\def\kv{\mbox{\boldmath $ k$}}

\def\rv{\mbox{\boldmath $r$}}
\def\cg{\check{g}}

\def\ha{\hat{a}}
\def\hb{\hat{b}}

\def\ab{(1+ \ha_+ \hb_-)^{-1}}
\def\ba{(1+ \hb_- \ha_+)^{-1}}

\title{Analytical Result on Electronic States around a Vortex Core in a Noncentrosymmetric Superconductor}

\author{\textsc{Yuki Nagai}$^{1}$\thanks{E-mail address: ss56079@mail.ecc.u-tokyo.ac.jp},
 \textsc{Yusuke Kato}$^{2}$ and  \textsc{Nobuhiko Hayashi}
$^{3}$}

\inst{$^{1}$Department of Physics, University of Tokyo, Tokyo 113-0033, Japan\\
$^{2}$Department of Basic Science, University of Tokyo, Tokyo 153-8902, Japan \\
$^{3}$Institut f$\ddot{u}$r Theoretische Physik, ETH-H$\ddot{o}$nggerberg, CH-8093 Z$\ddot{u}$rich, Switzerland }

\abst{We analytically study electronic bound states around a single vortex core 
in a noncentrosymmetric superconductor such as  ${\rm CePt_3Si }$ without mirror symmetry about the $ab$ plane. 
Considering a mixed spin-singlet-triplet Cooper pairing model, we obtain 
a formula about the local density of states (LDOS) around a vortex core in any direction of the magnetic field. 
The LDOS under a magnetic field perpendicular to the $c$ axis is quite different from 
that of typical s-wave or d-wave superconductors. 
From the ellipticity of the spatial pattern of the LDOS around a vortex core, one can experimentally estimate the pairing symmetry of ${\rm CePt_3Si}$, 
such as the position of the gap nodes and the ratio of the singlet component to the triplet component in the order parameter.}

\kword{CePt$_3$Si, Unconventional superconductivity, Vortex core, Local density of states,
Broken inversion symmetry, Quasiclassical theory}

\begin{document}
\maketitle
In a time reversal invariant system, the lack of inversion symmetry is connected with the presence of an antisymmetric spin-orbit coupling\cite{Gor'kov}.
Unconventional superconductivity in strongly correlated systems exists in heavy Fermion superconductors.
The discovery of the noncentrosymmetric heavy Fermion superconductor ${\rm CePt_3Si}$ \cite{Bauer} has attracted considerable attention.
Many theoretical and experimental studies on unconventional superconductivity without inversion symmetry
have been reported for a few years.\cite{Bauer,Bauer2,Kaur,frigeri1,frigeri_con,hayashi_con,hayashi_con2,hayashi_con3,Yip}
The vortex structure of the mixed state in this system was recently studied on the basis of the Ginzburg-Landau theory 
and the London theory.\cite{Kaur,Yip}
The vortex core structure was also studied by means of the quasiclassical theory of superconductivity, 
which enables us to calculate the physical quantities more microscopically.\cite{hayashi_con3}
However, no studies have been attempted to investigate the possible anisotropic spatial structure of the quasiparticle distribution around a vortex core in such a system.
In this paper, we analytically investigate the local density of states (LDOS) around a single vortex core in a noncentrosymmetric superconductor 
and devise a formula for LDOS that is applicable in any arbitrary direction of the magnetic field, thus 
revealing an anisotropic quasiparticle structure.
An anisotropic pairing symmetry is reflected by the real-space LDOS pattern
because of the presence of quasiparticles around a vortex core.\cite{ichioka,Schopohl2}
Therefore, 
for understanding the Cooper pairing without inversion symmetry, it is important to investigate the LDOS pattern.
The LDOS can be probed experimentally by STM \cite{Hess,Nishimori}.
If the LDOS pattern presented in this paper is consistent with that observed by STM, 
we can obtain information on the position of the gap nodes and the ratio of the singlet component to the triplet component in this material.
In other words; this can serve as one of the methods to experimentally investigate the pair potential directly in CePt$_3$Si.
Thus, we can also show that our theoretical formulation is capable of wide application.
 
We consider a mixed spin-singlet-triplet model to study the noncentrosymmetric superconductor.
Considering the results of numerical calculations by Hayashi \textit{et al.}\cite{hayashi_con3},
we assume that the spatial variations of the s-wave pairing component of the pair potential are the same as thos of the p-wave pairing component. 
$\hat{\Delta} =  \left[ \Psi \hat{\sigma}_0 + \Vec{d}_k \cdot \hat{\Vec{\sigma}} \right]i \hat{\sigma}_y 
 = \Delta(\rv) \left[ \tilde{\Psi} \hat{\sigma}_0 - \tilde{k}_y \hat{\sigma}_x + \tilde{k}_x \hat{\sigma}_y \right]i \hat{\sigma}_y $,
 where the s-wave pairing component $\Psi$, the $\Vec{d}$ vector $\Vec{d}_k = \Delta(- \tilde{k}_y, \tilde{k}_x,0)$, 
 the Pauli matrices in the spin space $\hat{\Vec{\sigma}}=(\hat{\sigma_x},\hat{\sigma}_y,\hat{\sigma}_z)$,  
 the unit matrix $\hat{\sigma}_0$, and the unit vector $\tilde{\kv}$ in the momentum space will be discussed later.
Here, the ratio of the singlet to the triplet component is defined in the bulk region as $\tilde{\Psi} = \Psi/\Delta$.
 This mixed s+p-wave model is proposed for ${\rm CePt_3Si}$\cite{frigeri_con}.
 
 In a system without inversion symmetry, there is a Rashba-type spin-orbit coupling with the form \cite{frigeri_con,hayashi_con2,frigeri1}
 \begin{equation}
 {\cal H}_1 = \sum_{\kv,\eta,\eta'} \alpha \Vec{g}_k \cdot \hat{\Vec{\sigma}}_{\eta \eta'} c^{\dagger}_{\kv \eta} c_{\kv \eta'},
 \end{equation}
where $\Vec{g}_k = \sqrt{\frac{3}{2}} \frac{1}{k_{\rm F}}(- k_y,k_x,0)$.
Here, $c^{\dagger}_{\kv \eta}$($c_{\kv \eta}$) is the creation (annihilation) operator for the quasiparticle state with
 momentum $\kv$ and spin $\eta$ and $\alpha$ ($>$0) denotes the strength of the spin-orbit coupling.
 We use units in which $\hbar = k_B =1$.
The $\Vec{g}_k$ vector, which is an antisymmetric vector ($\Vec{g}_{-k} = - \Vec{g}_k$), is parallel to the $\Vec{d}$ vector.\cite{frigeri1}

We calculate the LDOS around a vortex core on the basis of the quasiclassical theory of 
superconductivity\cite{Eilen,Serene,Larkin}.
We consider the quasiclassical Green function $\check{g}$ that has matrix elements in the Nambu (particle-hole) space as
\begin{eqnarray}
\check{g}(\rv,\tilde{\Vec{k}}, i \omega_n) =  
\left(
\begin{array}{cc}
\hat{g} &  \hat{f} \\
-  \hat{\bar{f}} &  \hat{\bar{g}}
\end{array}
\right)
,
\end{eqnarray}
where $\omega_n$ is the Matsubara frequency.
Throughout the paper, ^^ ^^ \textit{hat}" $\hat{a}$ denotes a $2 \times 2$ matrix in the spin space, and 
^^ ^^ \textit{check}" $\check{a}$ denotes a $4 \times 4$ matrix composed of the $2 \times 2$ Nambu space and the $2 \times 2$ spin space\cite{hayashi_con2}.

The Eilenberger equation, which includes the spin-orbit coupling term, is given as \cite{hayashi_con,hayashi_con2,Schopohl,Rieck,Choi}
\begin{equation}
i \Vec{v}_{\rm F}(\tilde{\kv}) \cdot \Vec{\nabla} \check{g} + 
[i \omega_n \check{\tau}_3 - \alpha \check{\Vec{g}_k} \cdot \check{\Vec{S}} - \check{\Delta},\check{g}]=0,
\end{equation}
where
\begin{eqnarray}
\check{\Vec{g}}_k = \left(
\begin{array}{cc}
\Vec{g}_k \hat{\sigma}_0 & 0  \\
0 & \Vec{g}_{-k} \hat{\sigma}_0
\end{array}
\right)
 = \left(
\begin{array}{cc}
\Vec{g}_k \hat{\sigma}_0 & 0 \\
0 & - \Vec{g}_k \hat{\sigma}_0
\end{array}
\right),\\
\Vec{g}_k = \sqrt{\frac{3}{2}}(- \tilde{k}_y, \tilde{k}_x,0), \: \: 
\check{\tau}_3 = 
\left(
\begin{array}{cc}
\hat{\sigma}_0 & 0 \\
0 & - \hat{\sigma}_0
\end{array}
\right),\\
\Check{\Vec{S}} = \left(
\begin{array}{cc}
\Vec{\hat{\sigma}} & 0 \\
0 & \Vec{\hat{\sigma}}^{\rm tr}
\end{array}
\right), 
\: \: \check{\Delta} = \left(
\begin{array}{cc}
0 & \hat{\Delta} \\
- \hat{\Delta}^{\dagger} & 0
\end{array}
\right).
\end{eqnarray}
Here, $\Vec{v}_{\rm F}(\tilde{\kv})$ is the Fermi velocity, 
$\hat{\Vec{\sigma}}^{\rm tr} = - \hat{\sigma}_y \hat{\Vec{\sigma}} \hat{\sigma}_y$, and the commutator
$[\check{a}, \check{b}] = \check{a} \check{b} - \check{b} \check{a}$.
$\check{g}$ should be subject to the normalization condition\cite{Eilen,Schopohl} $\check{g}^2 = \check{1}$,
where $\check{1}$ is a $4 \times 4$ unit matrix. 
We neglect the impurity effect and the vector potential because ${\rm CePt_3Si}$ is a clean extreme-type-II superconductor.\cite{Bauer}

The Eilenberger equation can be simplified by introducing a parametrization for the propagators that satisfy
the normalization condition.
Propagators are defined as $\check{P}_{\pm} = \frac{1}{2} \left( \check{1} \mp \cg \right)$, 
which were originally introduced in the studies of vortex dynamics \cite{eschrig}.
Using these propagators, we obtain the matrix Riccati equations as follows:
\begin{eqnarray}
 \Vec{v}_{\rm F} \cdot \Vec{\nabla} \ha_+ + 2  \omega_n \ha_+ + \ha_+ \hat{\Delta}^{\dagger} \ha_+   -
  \hat{\Delta} & & \nonumber \\ 
   + \:  i(\ha_+ \alpha (\Vec{g}_k \cdot \hat{\Vec{\sigma}})^{\rm tr} + \alpha 
 \Vec{g}_k \cdot \hat{\Vec{\sigma}} \ha_+) &=& 0,
 \label{eq:ra}\\
 \Vec{v}_{\rm F} \cdot \Vec{\nabla} \hb_- - 2  \omega_n  \hb_- - \hb_- \hat{\Delta} \hb_- + \hat{\Delta}^{\dagger} & & \nonumber 
 \\
  - \: i(\hb_- 
 \alpha 
 \Vec{g}_k \cdot \hat{\Vec{\sigma}} 
 +  \alpha (\Vec{g}_k \cdot \hat{\Vec{\sigma}})^{\rm tr} \hb_-)&=& 0,
 \label{eq:rb}
\end{eqnarray}
where
\begin{eqnarray}
\cg 
&=& -
\check{N}
\left(
\begin{array}{cc}
(\hat{1} - \ha_+ \hb_-) & 2i \ha_+ \\
- 2 i\hb_- & - (\hat{1} - \hb_- \ha_+)
\end{array}
\right),\\
\check{N} &=& \left(
\begin{array}{cc}
\ab & 0 \\
0 & \ba
\end{array}
\right).
\end{eqnarray}
We consider a single vortex along the $Z$ axis that tilts from the crystal $c$ axis by an angle $\phi$.
Now, we consider the $X$ axis on the $a$-$b$ plane.
We assume the spherical Fermi surface and consider the momentum vector $\kv = k(\cos \theta \sin \chi, \sin \theta \sin \chi, \cos \chi)$ 
in this coordinate system fixed to the magnetic field.
To obtain the quasiclassical Green functions, we consider the trajectories of the quasiparticle on the $X$-$Y$ plane \cite{ueno}. 
Because of a translational symmetry along the vortex, the trajectory of a quasiparticle with a momentum $|\kv_{\rm F}|$ and $k_Z \neq 0$ 
contributes to the quasiclassical Green functions in the same manner as that of a quasiparticle with a momentum $|\kv_{\rm F}| \sin \chi$  projected 
on the $k_X$-$k_Y$ plane.
In other words, the quasiclassical Green functions in a three-dimensional system can be converted into 
a set of quasiclassical Green functions in a two-dimensional system having a momentum with different amplitude.
Therefore, we determine the coordinates
\begin{eqnarray}
\left(
\begin{array}{c}
X  \\
Y 
\end{array}
\right)
=
\left(
\begin{array}{cc}
\cos \theta & - \sin \theta \\
\sin \theta & \cos \theta
\end{array}
\right)
\left(
\begin{array}{c}
x  \\
y 
\end{array}
\right)
\label{eq:real}
\end{eqnarray}
to solve eqs.~(\ref{eq:ra}) and (\ref{eq:rb}) along the direction 
of momentum in a two-dimensional system; the momuntum is in the $x$ direction.
Here, $y$ is the axis referred to as an impact parameter, $x$ is the axis perpendicular to $y$, and $r=\sqrt{x^2+y^2}=\sqrt{X^2+Y^2}$; 
the origin of these axes is at the vortex center.

Using the perturbative method developed by Kramer \textit{et al.} \cite{Kramer,ueno,eschrig2}, 
we can obtain the quasiclassical Green function around the vortex core in a low energy region ($|\omega_n| \ll |\Delta_{\infty} |$).
Here, $\Delta_{\infty}$ is a pair potential in the bulk region.
By expanding the matrix Riccati equations in the first order of $\alpha$, $y$, 
$|\Delta(r)|$, and $|\omega_n|$, we obtain the approximate solution as \cite{nagai}
\begin{eqnarray}
\hat{g} = \frac{\hat{O}_C e^{-2 (\sqrt{k_+ k_-} + \tilde{\Psi}) F(x) }}{2 F_D}   
 +\frac{\hat{O}_D e^{-2 | \sqrt{k_+ k_-} - \tilde{\Psi} | F(x) } }{2 F_C} , 
 \label{eq:green}
\end{eqnarray}
where
\begin{eqnarray}
F_{C,D} &=& \frac{1}{v_{\rm F}} \int_{-\infty}^{\infty} dx' \nonumber \\
& & \times \left[ -  \frac{ |\Delta(r')| }{|x'| }\frac{y}{\sin \chi} |\sqrt{k_+ k_-} \mp \tilde{\Psi}| 
	+ i \omega_n  \right]  \nonumber \\
    & &  \times \: e^{-2|\sqrt{k_+ k_-}  \mp \tilde{\Psi}|F(x')},\\
\hat{O}_{C,D} &=& \left(
\begin{array}{cc}
-i & \pm \sqrt{\frac{k_{-}}{k_{+}}}  \\
\mp \sqrt{\frac{k_{+}}{k_{-}}}  &-i 
\end{array}
\right).
\end{eqnarray}
Here,
$k_{\pm} = \tilde{k}_x \pm i \tilde{k}_y$ and $F(x) = \int_0^{|x|} d x' \Delta(x') /v_{\rm F}$.
The first and second terms in eq.~(\ref{eq:green}) are the Green functions referred to as $\hat{g}_{\rm I,II}$ on Fermi surfaces ${\rm I}$ and ${\rm II}$; 
this is because the spin orbit coupling splits the Fermi surface into two surfaces by lifting the spin degeneracy.\cite{frigeri1,frigeri_con}
Fermi ${\rm I}$ has a nodeless pair potential, and Fermi ${\rm II}$ has a pair potential with the line node (e.g., see Fig.~1 in ref.~\citen{hayashi_con2} ).

Near the vortex core ($|r| \ll \xi$), we use the following approximations,
\begin{eqnarray}
F_{C,D} &\sim&   -\frac{\tilde{y} | \sqrt{k_+ k_-} \mp \tilde{\Psi} | }{\sin \chi} +  \frac{i \tilde{\omega}_n}{|\sqrt{k_+ k_-} \mp \tilde{\Psi}|},
\end{eqnarray}
where
\begin{eqnarray}
F(x) = \frac{1}{v_{\rm F}} \int _0^{|x|} dx' \Delta(x') &\sim& 0, \\
\frac{1}{v_{\rm F}} \int_{- \infty}^{\infty} dx' e^{-2 a F(x')} 
&\sim& \frac{1}{a \Delta_{\infty}}, \\
\int_{0}^{\infty} dx' \frac{\Delta (x')}{x'} e^{-2 a F(x')} &\sim& \Delta_{\infty}.
\end{eqnarray}
Here, we use the dimensionless variables as $\tilde{\omega}_n = \omega_n / \Delta_{\infty}$ and $\tilde{y} = y/\xi_0$ with $\xi_0=v_{\rm F}/\Delta_{\infty}$.
Therefore, the LDOS around the vortex core is 
\begin{eqnarray}
\tilde{\nu}(\tilde{\rv},\tilde{\epsilon}) &=& 
- \Bigl \langle {\rm Re} \{ {\rm tr} \hat{g}^R(i \tilde{\omega}_n \to \tilde{\epsilon} + 0^+) \} \Bigr \rangle_{\tilde{k}} \\
&=& \tilde{\nu}_{\rm I}(0) \int_0^{\pi} \sin \chi d \chi \int_0^{2 \pi} \frac{d \theta}{4 } (\sqrt{k_+ k_-} + \tilde{\Psi}) \nonumber \\
& & \times \: \delta[h_{+}(\theta,\chi)]   \nonumber \\
& &  + \: \tilde{\nu}_{\rm II}(0) \int_0^{\pi} \sin \chi d \chi \int_0^{2 \pi} \frac{d \theta}{4 } \Bigl{|} \sqrt{k_+ k_-}
 - \tilde{\Psi} \Bigl{|} \nonumber \\
& & \times \: \delta[h_-(\theta,\chi)],
\label{eq:nu}
\end{eqnarray}
where
 $h_{\pm}(\theta,\chi) = \tilde{\epsilon} - \tilde{y}(\sqrt{k_+ k_-} \pm \tilde{\Psi})^2$$/$$\sin \chi$.
Here, $\tilde{\nu}_{\rm I,II}(0)$ denotes the density of states on Fermi surfaces I and II.
This integral reduces to an integral on the path satisfying $h_{\pm}(\theta,\chi) = 0$.
Therefore, at the points where $\tilde{\nu}(\tilde{\rv},\tilde{\epsilon})$ diverges, the following relations hold:
\begin{eqnarray}
h_{\pm} =
 \frac{\partial h_{\pm}}{\partial \theta} =
\frac{\partial h_{\pm}}{\partial \chi} = 0
   \label{eq:l3}
\end{eqnarray}
and the saddle point condition
\begin{equation}
\frac{\partial^2 h_{\pm}}{\partial \theta^2} \frac{\partial^2 h_{\pm}}{\partial \chi^2} -  \left( \frac{\partial^2 h_{\pm}}{\partial \theta \partial \chi } 
\right)^2 <0.
\label{eq:anten}
\end{equation}
These relations are derived from the following discussion.
Near the saddle point, 
$\tilde{\nu}(\tilde{\rv},\tilde{\epsilon})$ is rewritten as
\begin{eqnarray}
\tilde{\nu}(\tilde{\rv},\tilde{\epsilon}) &\sim& \int_{V \ni (0,0)} \delta(a x^2 + b y^2) dV,
\end{eqnarray}
with $a b <0$, and diverges logarithmically.

From eq.~(\ref{eq:l3}), it follows that 
\begin{eqnarray}
\tilde{x}_{\pm}&=& \frac{2 \tilde{\epsilon} \sin \chi \frac{\partial }{\partial \theta }\sqrt{\tilde{k}_x^2+ \tilde{k}_y^2}  }
{\left( \sqrt{\tilde{k}_x^2 + \tilde{k}_y^2 }
 \pm \tilde{\Psi} \right)^3}, 
 \label{eq:x}\\
\tilde{y}_{\pm}&=& \frac{\tilde{\epsilon} \sin \chi}{\left( \sqrt{\tilde{k}_x^2+ \tilde{k}_y^2 } \pm \tilde{\Psi} \right)^2},
\label{eq:y}\\
0 &=&  \frac{ \tilde{\epsilon} \cos \chi}{\sin \chi} - \frac{ \tilde{\epsilon} \frac{\partial }{\partial \chi}(\tilde{k}_x^2 + \tilde{k}_y^2) }
{\tilde{k}_x^2 + \tilde{k}_y^2  \pm \tilde{\Psi} \sqrt{\tilde{k}_x^2 + \tilde{k}_y^2 }}. \label{eq:0}
\end{eqnarray}
We can consider eq.~(\ref{eq:0}) as the equation to determine $\chi$.
The solution of eq.~(\ref{eq:l3}) is regarded as an enveloping curve of the trajectories of the quasiparticle 
 when $\chi$ is fixed.\cite{ueno}
Therefore, from eq.~(\ref{eq:real}), we can obtain the points where the LDOS diverges.

In momentum space, the relation between the axes fixed to the crystal axes ($k_x,k_y,k_z$) 
and those fixed to the magnetic field ($k_X,k_Y,k_Z$) is written as 
\begin{eqnarray}
k_x = k_X, \: \: \: \: 
 k_y = k_Y \cos \phi - k_Z \sin \phi, \\
k_z = k_Y \sin \phi + k_Z \cos \phi.
\end{eqnarray}
Here, we consider the rotation about the $a$ axis; this rotation reveals that the $X$ axis is equal to the $a$ axis.
Therefore, using the spherical coodinates fixed to the magnetic field
($k_X = k_{\rm F} \cos \theta \sin \chi$, $k_Y = k_{\rm F} \sin \theta \sin \chi$, $k_Z = k_{\rm F} \cos \chi$), we obtain 
\begin{eqnarray}
\tilde{k}_x &=& \cos \theta \sin \chi, \\ 
\tilde{k}_y &=& \sin \theta \sin \chi \cos \phi - \cos \chi \sin \phi.
\end{eqnarray}

First, we consider the system in a magnetic field parallel to the $c$ axis ($\phi = 0$).
Equation (\ref{eq:0}) leads to $\chi = \frac{\pi}{2}$. 
Therefore, we obtain the parametric equations
\begin{eqnarray}
\tilde{X}_{\pm}= - \frac{\tilde{\epsilon} \sin \theta}{(1 \pm \tilde{\Psi})^2} \: \: {\rm and} \: \: \:\tilde{Y}_{\pm} =\frac{\tilde{\epsilon} \cos \theta}{(1 \pm \tilde{\Psi})^2}
\end{eqnarray}
 from eqs.~(\ref{eq:anten}), (\ref{eq:x}), and (\ref{eq:y}).
This result is consistent with the numerical calculation by Hayashi \textit{et al}.\cite{hayashi_con3} because 
this spatial LDOS pattern is in the form of two concentric circles.
This result shows that the quasiparticles rotate around a vortex core, and the LDOS exhibits the two-gap property.
The ratio of the singlet component to the triplet component, $\tilde{\Psi}$, determines the ratio of the radius between the two concentric circles.

\begin{figure}[htbp]
 \begin{center}
  \includegraphics[width=7cm,keepaspectratio]{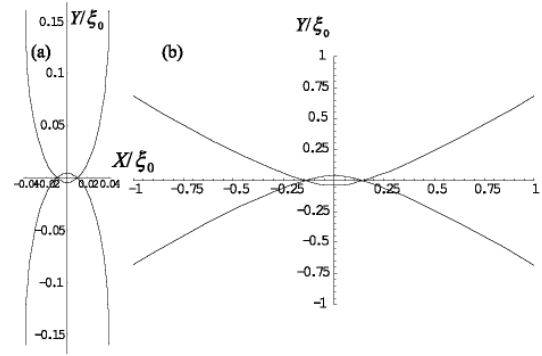}
\caption{Distribution of the points where the local density of stetes diverges around the vortex perpendicular to the $c$ axis. 
(a) Fermi I ($h_+$) and (b) Fermi II ($h_-$). Here, $\tilde{\Psi} = 0.5$ and $\epsilon/\Delta_{\infty} = 0.01.$}
\label{fig:hp}
 \end{center}
\end{figure}

\begin{figure}[htbp]
 \begin{center}
  \includegraphics[width=8cm,keepaspectratio]{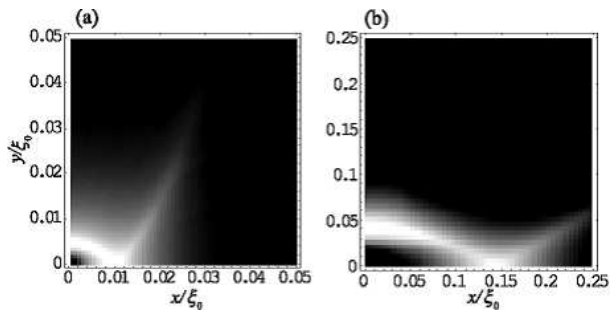}
\caption{Distribution of the LDOS around the vortex perpendicular to the $c$ axis. (a) Fermi I ($h_+$) and (b) Fermi II ($h_-$).
Here, $\tilde{\Psi} = 0.5$ and $\epsilon/\Delta_{\infty} = 0.01.$ Notice the difference in spatial extent between these LDOS patterns.}
\label{fig:3ds}
 \end{center}
\end{figure}

Second, we consider the system in a magnetic field perpendicular to the $c$ axis ($\phi = \frac{\pi}{2}$).
In this case, $\chi = \frac{\pi}{2}$ follows from eq.~(\ref{eq:0}).
Therefore, we obtain
 the following parametric equations:
\begin{eqnarray}
\tilde{X}_{\pm} = \frac{- \tilde{\epsilon} \sin \theta}{(|\cos \theta| \pm \tilde{\Psi})^3}\left\{
3 | \cos \theta| \pm \tilde{\Psi}
 \right\},  
 \label{eq:xperp}\\
\tilde{Y}_{\pm} = -2 \:  \frac{\tilde{\epsilon}  \sin^2 \theta {\rm sign}[\cos \theta] }{(|\cos \theta| \pm \tilde{\Psi})^3} 
+ \frac{ \tilde{\epsilon} \cos \theta }{(|\cos \theta| \pm \tilde{\Psi})^2}, 
\label{eq:yperp}
\end{eqnarray}
where $\tilde{\Psi} < |\cos \theta| \: ({\rm \: in \: the \: case \: of \: } h_-)$ 
using eqs.~(\ref{eq:anten}), (\ref{eq:x}), and (\ref{eq:y}) with regard to Fermi I ($h_+$) and II ($h_-$).
In Fig.~\ref{fig:hp}, we show these LDOS patterns around the vortex core for a fixed energy ${\tilde \epsilon}$.
The pattern about Fermi I in Fig.~\ref{fig:hp}(a) can be regarded as a result of the shape of the order parameter, such as 
two circles with a shift of the centers (shown in Fig.~1(a) in ref.~\citen{hayashi_con2}).
If the relation ${\tilde \Psi} < |\cos \theta |$ between $\tilde{\Psi}$ and $\theta$ does not exist in the case of $h_-$,
the pattern about Fermi II is similar to that in a d-wave superconductor\cite{Schopohl2,ichioka} 
because of the presence of four nodes in the pair potential on
the circular line $\chi =\frac{\pi}{2}$ on the Fermi surface.
However, the actual pattern about Fermi II (Fig.~\ref{fig:hp}(b)) is similar to \textit{a part of} that in a d-wave superconductor
because of the presence of this relation between ${\tilde \Psi}$ and $\theta$ originating from the three-dimensional anisotropic pair potential.
These results show that the quasiparticles on Fermi I are bound around a vortex, 
while those on Fermi II move in the direction associated with one node to the direction associated with the other node.
We also calculate the distribution of the LDOS from eq.~(\ref{eq:nu}) by numerical integration (Fig.~\ref{fig:3ds}).
The LDOS in Fig.~\ref{fig:hp} is consistent with that in Fig.~\ref{fig:3ds}.
In the limit $\tilde{\Psi} \rightarrow 0$, the LDOS patterns in a magnetic field perpendicular to the $c$ axis 
become rather isotropic because the line nodes in a gap disappear.
Therefore, the strongly anisotropic LDOS patterns, as shown in Fig.~\ref{fig:3ds}(b), suggest 
that the triplet channel and the singlet channel are mixed.

\begin{figure}[htbp]
 \begin{center}
  \includegraphics[width=5.75cm,keepaspectratio]{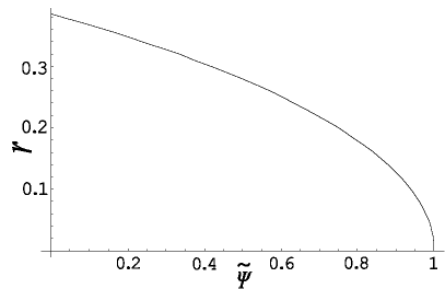}
\caption{Relation between the ellipticity $r$ of the pattern around a vortex in the LDOS pattern about Fermi II and 
the ratio of the singlet component to the triplet component, $\tilde{\Psi}$.}
\label{fig:ratio}
 \end{center}
\end{figure}

We can estimate the ratio of the singlet component to the triplet component, $\tilde{\Psi}$, from the spatial LDOS pattern shown in Fig.~\ref{fig:3ds}(b).
From eqs.~(\ref{eq:xperp}) and (\ref{eq:yperp}),
we calculate the ratio of the intercept $X_0$ on the $X$ axis to the intercept $Y_0$ on the $Y$ axis in the LDOS pattern for Fermi II.
The ratio between these intercepts  $r$ is written as
\begin{equation}
r =  \frac{ (-5 \tilde{\Psi} + \sqrt{24 + \tilde{\Psi}^2})^3}
{108(1 - \tilde{\Psi})^2 (-\tilde{\Psi} + \sqrt{\tilde{\Psi}^2 + 24})\sqrt{1 - \left( \frac{\tilde{\Psi} + \sqrt{\tilde{\Psi}^2 +24}}{6} \right)^2}}
, 
\end{equation}
where $r = Y_0/X_0 $ (Fig.~\ref{fig:ratio}); this ratio does not depend on the energy $\tilde{\epsilon}$.
In other words, we can obtain the ratio $\tilde{\Psi}$ from the ellipticity $r$ of the shape around a vortex.
We can also obtain the position of the gap-node $\theta_{\rm node}$ with the relation $|\cos \theta_{\rm node}| = \tilde{\Psi}$.
Now, CePt$_3$Si has a tetragonal crystal structure.
We have also investigated how the present result is influenced by an anisotropic mass tensor; thus, we found 
that the LDOS pattern is not significantly affected by the ellipsoidal Fermi surface.


In conclusion, we investigated the local density of states around a vortex core in a noncentrosymmetric superconductor.
We derived an analytical formula of the LDOS in any direction of the magnetic field.
In a magnetic field parallel to the $c$ axis, we found that the LDOS pattern is in the form of two concentric circles; 
this is consistent with the numerical calculation by Hayashi \textit{et al.} \cite{hayashi_con3}.
In a magnetic field perpendicular to the $c$ axis, we found that the LDOS pattern about Fermi I is as shown in Fig.~\ref{fig:hp}(a) 
and 
that about Fermi II extends far from the vortex center
as shown in Figs.~\ref{fig:hp}(b) and \ref{fig:3ds}(b); this pattern is similar to that of a d-wave superconductor,
but it is different from a four-fold d-wave superconductor.
The anisotropic LDOS patterns indicate the mixed singlet-triplet channels.
We can obtain the ratio of the singlet component to the triplet component in ${\rm CePt_3Si}$
from the ellipticity of the shape around a vortex as shown in Fig.~\ref{fig:hp}(b).
While we found that the effect of the deformation from the spherical
to the ellipsoidal Fermi surface is not quite significant,
it may be important in the future to investigate a more realistic band structure.
Our analytical formulation presented in this paper
is advantageous because it can be easily generalized to a system
with anisotropic Fermi surfaces.





\section*{Acknowledgment}
We thank D. F. Agterberg and M. Sigrist for helpful discussions.
This work is supported by a Grant-in-Aid for Scientific Research (C)(2)
No.17540314 from JSPS.
One of us (N.H.) is supported by the 2003 JSPS Postdoctoral Fellowships for Research Abroad.




\begin{thebibliography}{99}
\bibitem{nagai} The details of the calculation to obtain this solution will be reported later in a full length paper.
\end{thebibliography}
\end{document}